**Comment on 'Straight lightning as a signature of macroscopic dark matter By Nathaniel Starkman, Harrison Winch, Jagjit Singh Sidhu, and Glenn Starkman, Phys. Rev. D 103, 063024 – Published 18 March 2021'**

**Vernon Cooray** [1,*], **Gerald Cooray** [2], **Marcos Rubinstein** [3] and **Farhad Rachidi** [4]

[1] Department of Electrical Engineering, Uppsala University, 752 37 Uppsala, Sweden.
[2] Karolinska Institute, Stockholm, 171 77 Solna, Sweden.
[3] HEIG-VD, University of Applied Sciences and Arts Western Switzerland, 1401 Yverdon-les-Bains, Switzerland
[4] Electromagnetic Compatibility Laboratory, Swiss Federal Institute of Technology (EPFL), 1015 Lausanne, Switzerland

In the discussed paper, the authors have made several assumptions and statements concerning the initiation of lightning flashes by macroscopic dark matter passing through the atmosphere. The authors suggest that the path of dark matter can be identified by looking for lightning with straight channels. Even though we agree with the suggestion of the authors that macroscopic dark matter can give rise to straight lightning channels, there are several statements in the paper that are not clear and which could lead to misinterpretation of the results gathered in any future search for such lightning channels. Our comments on the paper are the following:

1. In the discussed paper, the authors assumed that once a plasma channel resulting from a macroscopic dark matter is created at an appropriate location, a lightning leader may be 'locked in' to the plasma channel and propagate downwards along it to the ground. A possible interpretation of the 'lock in' process is that an existing lightning discharge intercepts the macroscopic dark matter generated plasma channel or vice versa. We will make reference to this interpretation in the next paragraph. However, the authors may have meant something else since they do not provide details on what specifically is meant by the 'lock in' mechanism.

We would like to point out that as the plasma channel is created in the high electric field of the thundercloud, electric charges will be continuously induced at its forward-moving tip [1]. As the channel of the macro passes through the cloud, there are several ways that it can interact with the charged cloud. Let us consider the case where the speed of the macro is faster than the speed of propagation of electrical discharges in virgin air. As the macro passes through the cloud, there is a chance that the macro channel could intercept an already existing electrical discharge inside the cloud (see our interpretation above), but this event is highly improbable. A more probable scenario would be the following. Since the macro channel is highly conducting, it will be polarized when exposed to an electric field. As the macro channel passes through the cloud and when it is in the vicinity of a charge center, the channel will be polarized and charge of opposite polarity to that of the charge center will be induced on the region of the macro channel which is in the vicinity of the charge center. If the background electric field is large enough, the induced charge could, in turn, be high enough to generate an electrical discharge in the form of a streamer-assisted leader traveling from the macro channel into the charge center. Such discharges would be similar to that of the preliminary breakdown activity taking place during the initiation of lightning flashes. In this case, the macro channel will be physically connected to the charge center. This connection could further enhance electrical breakdowns inside the charge center, similar to the case of an upward leader from triggered lightning or from a tall structure that enters into the cloud. Different forms of electrical discharges that drive other lightning processes, such as dart leaders, continuing currents and M-components along such leaders could also take place when the macro channel is connected to the charge center and all these discharges could generate ionization waves along the macro channel feeding the charge accumulated on it. However, since the macro channel is moving faster than the electrical discharges propagating in virgin air (such as stepped leaders), this charging process will not be able to generate electrical discharges that shoot out from the tip of the macro channel, keeping therefore the channel straight and without branches. This electrical activity taking place along the macro channel would keep its conductivity high and prevent the channel from decaying prematurely. Of course, the situation would be identical to that described above even if the macro channel intercepted an already existing electrical activity inside the cloud.

As the macro channel propagates through the cloud, electrical breakdown could first take place to the positive charge center and, as the macro channel continues its downward journey, it could give rise to an electrical discharge also towards the negative charge center. In this case, two charge centers will be connected through the macro channel, which could trigger a cloud discharge, namely, electrical discharges propagating between the two charge centers assisted by the macro channel. However, since the macro still continues to propagate downwards, electrical discharges either from the positive charge center or the negative charge center or from both could feed the macro channel depositing charge along it. This is similar to a lightning flash that starts as a cloud flash and then culminates in a ground flash. The polarity of the charge on the macro channel when it reaches the ground will be determined by the amount of charge in the charge centers and the intensity of the electrical activity taking place inside individual charge centers.

Let us now consider the case where the macro is moving slower than the propagation of electrical discharges in virgin air. For example, electrical discharges in virgin air first take place in the form of streamer discharges. The speed of these streamers could be of the order of $10^6$ m/s [2] and, if the speed of the macroscopic dark matter particle is lower than that of the streamers, then these streamers will move ahead of the plasma channel generating leader type discharges from the tip of the macro channel. As the macro passes close to the charge centers of the cloud, such electrical discharges could shoot out in the form of stepped leaders from the tip of the macro channel towards charge centers when a sufficient amount of charge is accumulated at the tip due to the polarization of the macro channel in the electric field created by the charge centers. Such discharges could create a connection between the macro channel and the charge centers in the cloud. Since the macro is propagating slower than these electrical discharges, the latter propagate ahead of the tip of the macro channel. However, the path of these electrical discharges will be determined by the background electric field and may not coincide with the subsequent path of the macro channel. Thus, some parts of the discharge channel could remain straight while the rest become tortuous. Probably, it could lead to a straight channel with tortuous branches. Since the stepped leader is moving faster than the macro, the former will reach the ground first and the tip of the macro channel later. In this case, one might have a normal looking tortuous lightning channel to ground followed in time (the time interval being dependent on the speed of the macro) by a straight channel, both fed by the same electrical activity in the cloud. During the return stroke, both the tortuous channel including branches and the straight channel would be visible to the naked eye.

Interestingly, the propagation of a discharge ahead of the tip of a macro channel is similar to that taking place during rocket-and-wire triggered lightning flashes in which, as the wire extends upwards, charges accumulate along the wire due to the background electric field. When the charge at the tip of the wire becomes sufficiently large, an upward leader will start due to ionization processes taking place ahead of the tip. The breaking out of a leader-like discharge from the tip of the macro channel could take place exactly in the same manner.

2. The authors claim that there would not be sufficient time for the upward connecting leaders to be initiated from ground towards the down-coming plasma channel. However, experimental data [3] suggest the existence of upward connecting leaders with dart leaders travelling towards ground at speeds exceeding $10^6 - 10^7$ m/s. This indicates that the charged plasma channel moving downwards with a speed of 2.5 x $10^5$ m/s as considered in the paper will provide ample time for the generation of an upward connecting leader.